\documentclass[journal]{IEEEtran}
\pdfoutput=1
\usepackage{amsmath}
\usepackage{amssymb,amsfonts,textcomp}
\usepackage{array}
\usepackage{supertabular}
\usepackage{hhline}
\usepackage[export]{adjustbox}
\usepackage{graphicx}
\usepackage{epstopdf}
\usepackage{caption}
\usepackage{float}
\usepackage{subcaption}
\usepackage[load=prefix,load=named]{siunitx}
\makeatletter
\newcommand\arraybslash{\let\\\@arraycr}
\makeatother
\renewcommand\arraystretch{1.3}

\ifCLASSINFOpdf
\else
\fi

\begin{document}
%
\title{Space qualified nanosatellite electronics platform for photon pair
experiments}
%
%
%
\author{Cliff~Cheng,~Rakhitha~Chandrasekara,~Yue Chuan~Tan, and~Alexander~Ling\\
Centre for Quantum Technologies, National University of Singapore,\\
Block S15, 3 Science Drive 2,\\
 Singapore. 117543}
\maketitle

\begin{abstract}
We report the design and implementation of a complete electronics platform for
conducting a quantum optics experiment that will be operated on board a 1U
CubeSat (a 10 x 10 x 10 cm satellite). The quantum optics experiment is designed
to produce polarization-entangled photon pairs using non-linear optical
crystals and requires opto-electronic components such as a pump laser, single
photon detectors and liquid crystal based polarization rotators in addition to
passive optical elements.
The platform provides mechanical support for the optical assembly. It also
communicates autonomously with the host satellite to provide experiment data for
transmission to a ground station. A limited number of commands can be
transmitted from ground to the platform enabling it to switch experimental
modes. This platform requires less than 1.5W for all operations, and is space
qualified. The implementation of this electronics platform is a major step on
the road to operating quantum communication experiments using nanosatellites.
\end{abstract}

\begin{IEEEkeywords}
photon entanglement, space-based quantum communication, CubeSat, nanosatellite
\end{IEEEkeywords}

%

\section{Introduction}
%
%
%
%
%
%

A number of proposals \cite{Renner2009_1,TScheidl2013,Jennewein2014} have been been published
for building global quantum communication networks using satellites that host
quantum light sources or detectors. Efforts are underway to implement the first
demonstrations. Together with collaborators \cite{Morong2012a,Daniell2012}, we
have proposed that nanosatellites (spacecraft that have a mass below \SI{10}{\kilo\gram}) have
a role to play in this effort. They could act as demonstrators to raise the
technology readiness level of essential components and also as the final
platforms that transmit and receive single photons from ground-based stations or
other satellites. In particular, we propose that nanosatellites can effectively
host robust and compact sources of polarization-entangled photon pairs, which
are the workhorse for entanglement-based quantum communication. The decreasing
cost of launching a nanosatellite into low earth orbit has added impetus to this
approach \cite{Coopersmith2011}.

In order to use nanosatellites effectively, we are working to create small,
low-resource and rugged photon pair sources that
are fully compatible with the popular CubeSat standard \cite{Woellert2011b}. The
photon pair source that we are building is called the Small Photon-Entangling
Quantum System (SPEQS), and it is an integrated instrument combining low-power
electronics and a rugged optical assembly. 

The SPEQS instrument is designed to produce and detect pairs of photons via a process known
as spontaneous parametric down conversion (SPDC) \cite{Burnham}. In the SPEQS design
a \SI{405}{\nano\meter} pump beam interacts with a nonlinear optical crystal. With
some probability a pump photon is converted into a pair of daughter photons
obeying energy and momentum conservation. The daughter photons are strongly
correlated in polarization. Consequently, a measurement of the polarization
correlation is a good mechanism for monitoring the performance of the entangled
photon source.  The aim of the first SPEQS instrument is to demonstrate that the precisely aligned SPDC
source survives launch and can perform reliably in low Earth orbit. This
performance will be monitored by measuring the quality of the polarization
correlations. 

The electronics platform for the SPEQS instrument must operate a number of
opto-electronic devices efficiently. These include the diode laser for the pump beam,
the Geiger-mode avalanche photodiodes (GM-APD) for detecting the downconverted photons and
polarization rotators. In addition to the operation of the opto-electronic devices, the
SPEQS instrument must store experiment data. Experiment data are primarily in
the form of photo-detection events generated by the GM-APDs, and associated
house-keeping data such as laser power and temperature. Data must be stored on the SPEQS instrument
 before transfer to the spacecraft bus for transmission to ground stations. 

The platform also serves as the mechanical interface between the spacecraft and the optical
assembly. In this paper, we report the design and implementation of the
electronics platform that enables the SPEQS instrument to operate autonomously
on board a 1U CubeSat. 

\section{ Main modules of the electronics platform} 

The electronics platform is designed around the Cypress CY8C3666
Programmable-System-On-Chip (PSoC3) microcontroller. The PSoC3 is widely used in
white goods and is easily available as a commercial-off-the-shelf (COTS)
component. The PSoC3 is essentially an 8-bit 8051 microcontroller bundled
together with many digital components such as counters, timers,
analogue-to-digital converters (ADC), digital-to-analogue converters (DAC) and
pulse-width-modulation (PWM) devices. These active components are widely used
for signal preparation and conditioning in quantum optics experiments, and it is
convenient to access all these devices on a single chip. The functional blocks
are configured using a development environment supplied by Cypress (PSoC3
Creator Integrated Development Environment). 

With these available building block components, there is less need for external
chips or glue-logic circuitry reducing the physical footprint of the platform.
However, to enable complex analogue signal flows there is still the need for
additional circuitry composed of an assortment of switch capacitors, op-amps,
comparators, and digital filter blocks. 

\begin{figure}[h]
\begin{center}
\includegraphics[width=0.395\textwidth,keepaspectratio]{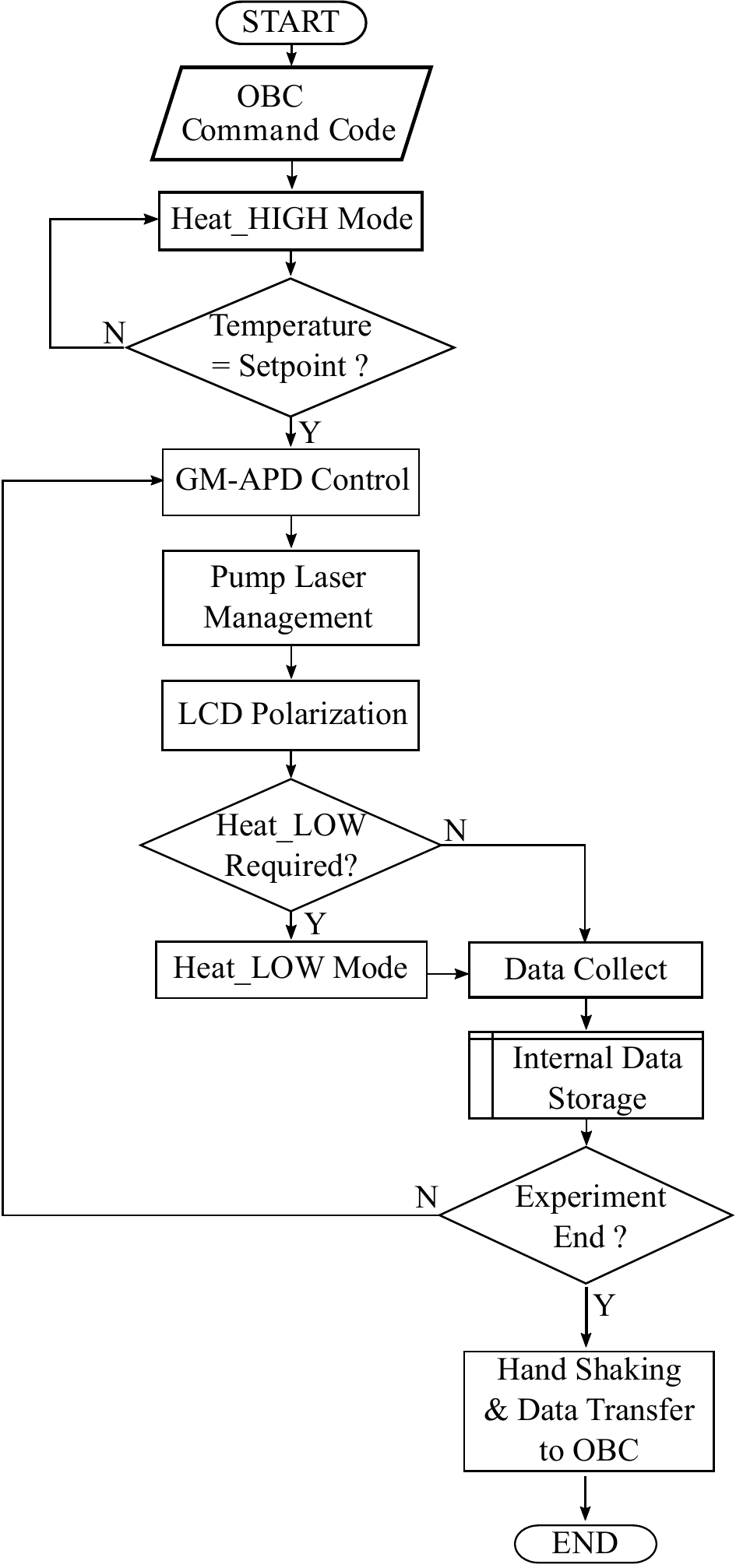}
\caption{ Concept of operations for the SPEQS electronics platform.}
\label{fig:SPEQSElectronicsv121B1-img001}
\end{center}
\end{figure}

FIG.~\ref{fig:SPEQSElectronicsv121B1-img001} illustrates the concept of
operations (ConOps) after the platform is powered on by the satellite's on-board
computer (OBC). Upon activation, the platform receives a command code from the
OBC that determines the experimental profile for the optical experiment. A
heating mode is activated to bring the optical experiment within an acceptable
temperature range. When the temperature range is achieved, the main
opto-electronics components (GM-APD, pump laser, liquid crystal polarization
rotator) are turned on sequentially. During the experiment, the heating mode is
maintained. Upon conclusion of the experiment, data is stored on a memory module
before transfer to the OBC. 

\begin{figure}[h]
\begin{center}
\includegraphics[width=0.52\textwidth,keepaspectratio]{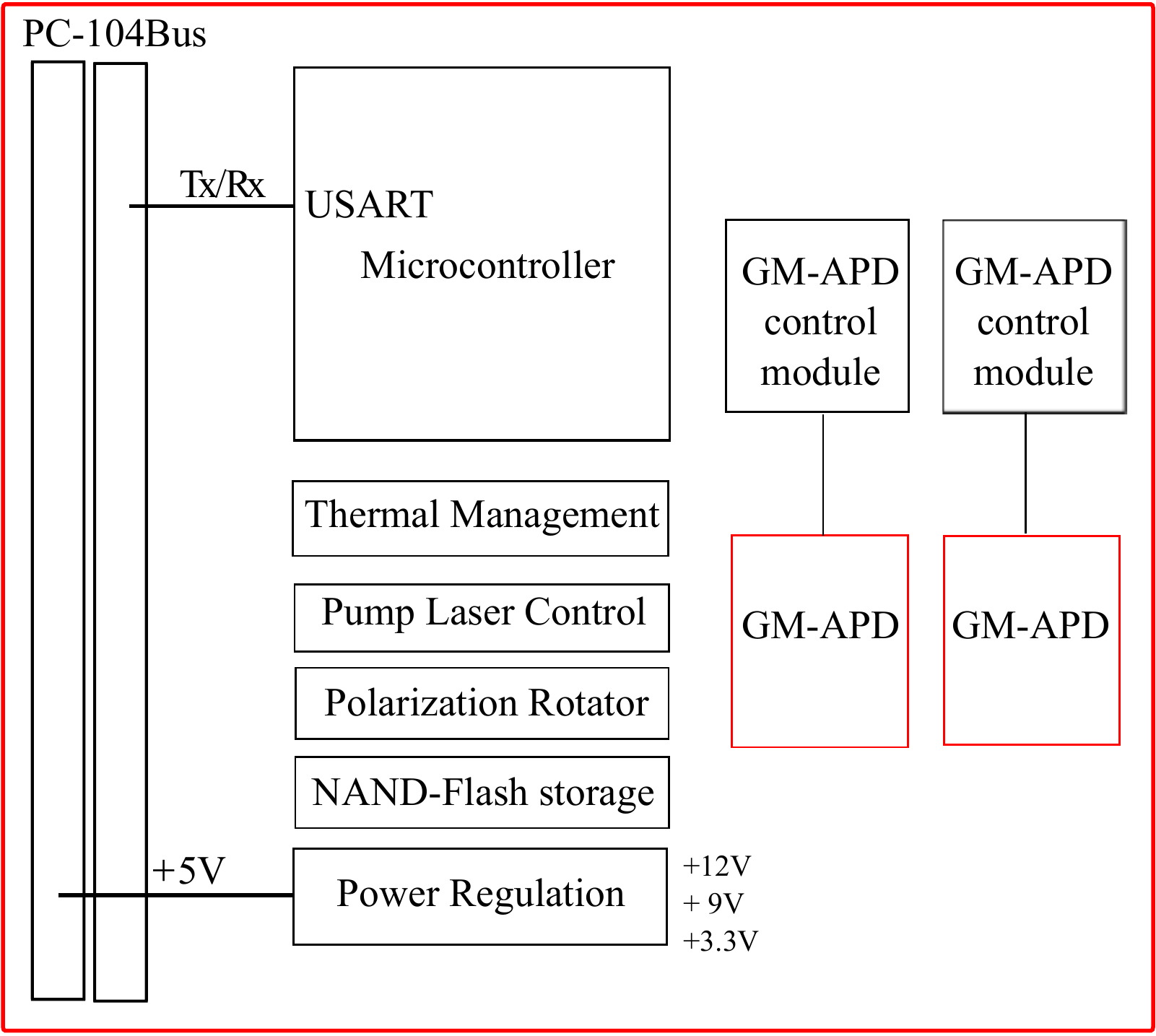}
\caption{Main modules of the electronics platform are: thermal management, pump
laser control, polarization rotator control, data storage, power regulation and
GM-APD control. A communication module (USART) allows asynchronous data
communication from the electronics platform to the satellite's OBC. The platform
incorporates a PC-104 bus to be compatible with the CubeSat standard. A microcontroller
(Cypress PSOC3) coordinates all modules on the platform.}
\label{fig:SPEQSElectronicsv121B1-img002}
\end{center}
\end{figure}

The main functional modules of the platform are shown in
FIG.~\ref{fig:SPEQSElectronicsv121B1-img002} With the exception of the power
regulation module (implemented using COTS regulators), these modules will be
described in the rest of this section. The GM-APD and polarization rotator
modules will be presented in more detail, as they are relatively complex mixed
digital-analogue circuits.

\subsection{GM-APD module}

The GM-APD (Laser Components SAP500) used in the SPEQS instrument is a
reach-through device with a large active area and relatively high detection
efficiency whose breakdown voltage ($V_{br}$) is slightly above \SI{120}{\volt} at room
temperature. Under normal operating conditions the bias voltage ($V_{bias}$)
applied to a GM-APD is in excess of $V_{br}$, and the photon detection
efficiency is a function of this excess voltage ($V_{E}$ = $V_{bias}$ -
$V_{br}$). When a photo-electron is present in the active area of the GM-APD an
electron avalanche is generated and detected as a current pulse. From the pulse
rate, the brightness of the entangled photon source is determined. Pairs of
photons are identified via correlated pulses from two GM-APDs.

\begin{figure*}[ht]
  \begin{minipage}[c]{0.48\textwidth}
 \centering
   \includegraphics[scale=0.58]{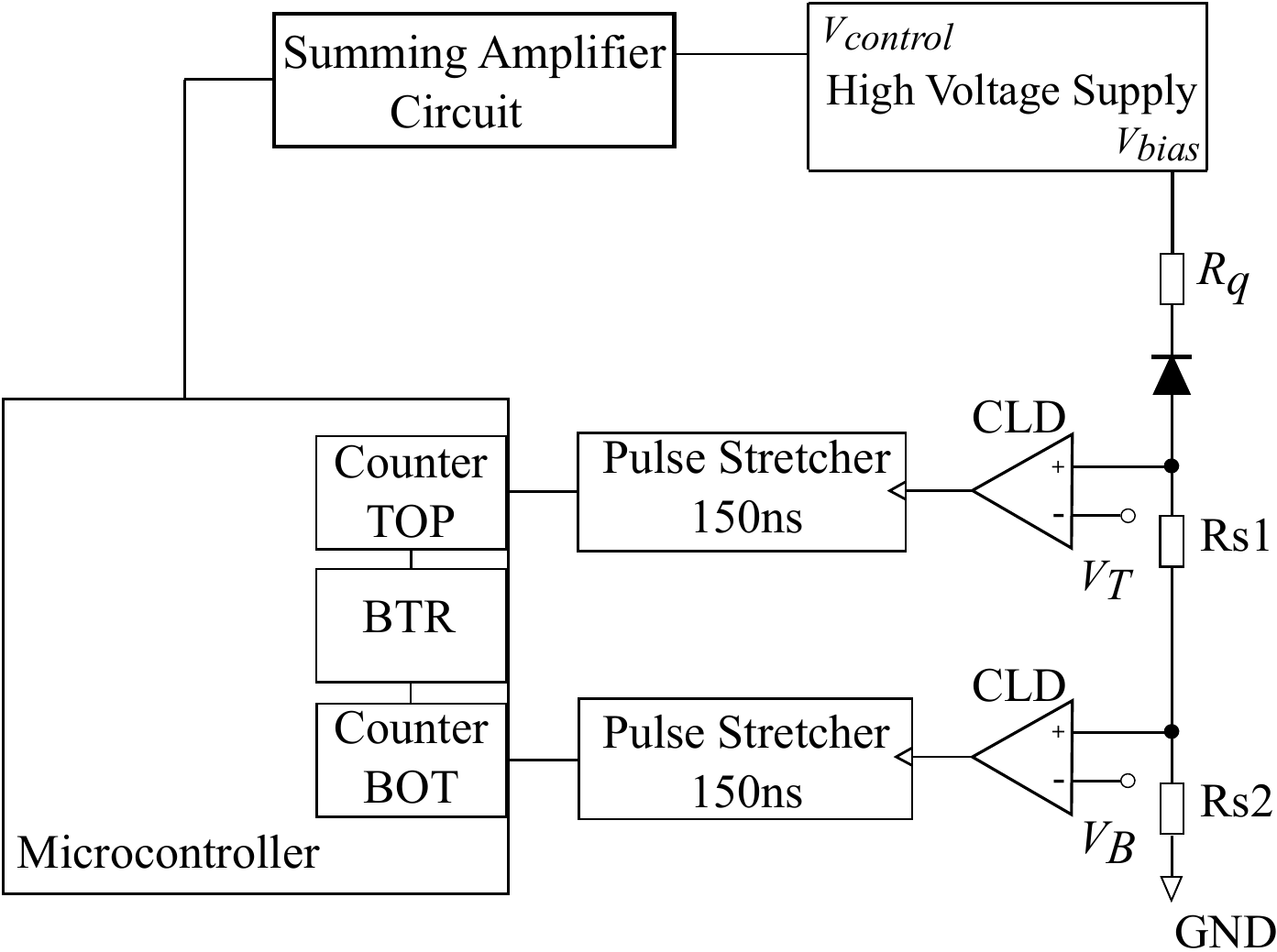}
  \subcaption{{}\label{}}
  \label{fig:SPEQSElectronicsv121B1-img003}
  \end{minipage}
\hfil
 \begin{minipage}[c]{0.48\textwidth}
 \centering
  \vspace{1.1cm}
  \includegraphics[scale=0.52]{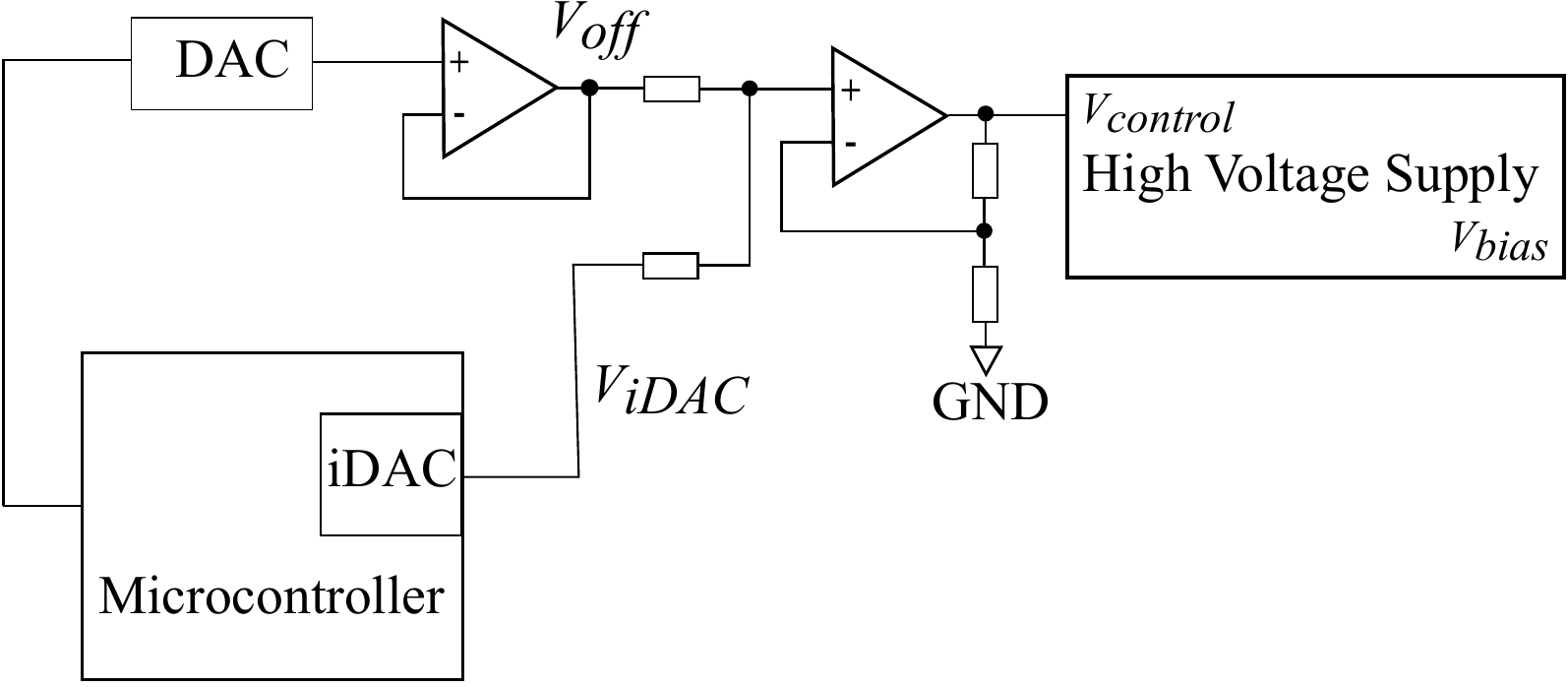}
  \vspace{1.32cm}
  \subcaption{{}\label{}}
  \label{fig:SPEQSElectronicsv121B1-img003}
  \end{minipage}
  \caption{ (a) The GM-APD module. The avalanche pulse is detected by a pair of sense
resistors that produce two pulses ({\textquotedblleft}top{\textquotedblright}
and {\textquotedblleft}bottom{\textquotedblright} that have two different pulse
heights. The {\textquotedblleft}top{\textquotedblright} resistor is Rs1 while
the {\textquotedblleft}bottom{\textquotedblright} resistor is Rs2. Each pulse is
compared against a constant level discriminator (CLD) to ensure that the
bottom-to-top ratio (BTR) of pulses is maintained within a certain range. Each CLD has an individual reference voltage ($V_{T},V_{B}$). When
this range is exceeded, it is taken to mean that the bias voltage should be
reduced. When the ratio falls below the range, the bias voltage is increased.
This (window comparator) technique enables the photo-detection event to
contribute to the GM-APD circuit and does not rely on temperature measurements.
   (b) The summing amplifier circuit. The output of current digital-to-analog
converter (iDAC) embedded in the PSoC3 is converted into a range of voltages
($V_{iDAC}$). This is added to an offset voltage ($V_{off}$) to produce the
control to the high voltage supply ($V_{control}$). It is necessary to use this
``summing amplifier'' as range for $V_{control}$ is outside the range of the DAC
components within the microcontroller.}
\label{fig:SPEQSElectronicsv121B1-img003}
\end{figure*}

FIG.~\ref{fig:SPEQSElectronicsv121B1-img003}(a) shows the block diagram of the
GM-APD module which incorporates a pulse detection method that uses a real-time
feedback control loop. The control loop maintains a fixed $V_{E}$ in order for
the detection efficiency to be constant over a range of operating temperatures
($V_{br}$ changes with temperature). 

The GM-APD is passively quenched which requires the current during an avalanche
event to be below \SI{50}{\micro\ampere} (the latch current). The current value is limited by
using an adequately large quench resistor, $R_{q}$, such that the value of
$V_{E}$/$R_{q}$ is smaller than the latch current of the given GM-APD. During
quench, the bias voltage across the GM-APD falls below $V_{br}$. The bias
voltage then recovers to its nominal value exponentially with a time constant $t$
($t  = R_{q}$$C$ where $C$ is the inherent capacitance of the GM-APD plus
parasitic capacitance). A typical parasitic capacitance is on the order of \SI{1}{\pico\farad},
while the inherent capacitance of a GM-APD is about \SI{3.3}{\pico\farad} putting the recovery
time constant at approximately \SI{2.4} {\micro\second}.


For each GM-APD,  there is a pair of {\textquotedblleft}top{\textquotedblright}
and {\textquotedblleft}bottom{\textquotedblright} sense resistors. The
{\textquotedblleft}bottom{\textquotedblright} sense resistor value is selected
so that it produces an avalanche pulse whose peak voltage is half in value to
that of the {\textquotedblleft}top{\textquotedblright} sense resistor. Each
pulse is compared against a constant level discriminator (CLD) to ensure that
the ratio of bottom-to-top (BTR) pulses is maintained within a pre-calibrated
range. When this range is exceeded, it is taken to mean that the bias voltage
should be reduced. When the ratio falls below the range, the bias voltage is
increased. This technique enables avalanche events to contribute to the bias
voltage control and does not rely on temperature measurements.

A pulse-stretching circuit converts the output of the CLDs to be \SI{150}{\nano\second} in
duration, in order for the microcontroller to detect the pulses. Within the 
microcontroller, a 16-bit counter is configured to
register and accumulate electronic pulses from each CLD.  The counter
accumulates its register value into a software variable every \SI{50}{\milli\second}. Every
second this variable is saved into a flash memory device as part of the
experiment record.

\begin{figure*}[ht]
  \begin{minipage}[c]{0.48\textwidth}
 \centering
 \vspace{1.3cm}
   \includegraphics[scale=0.455]{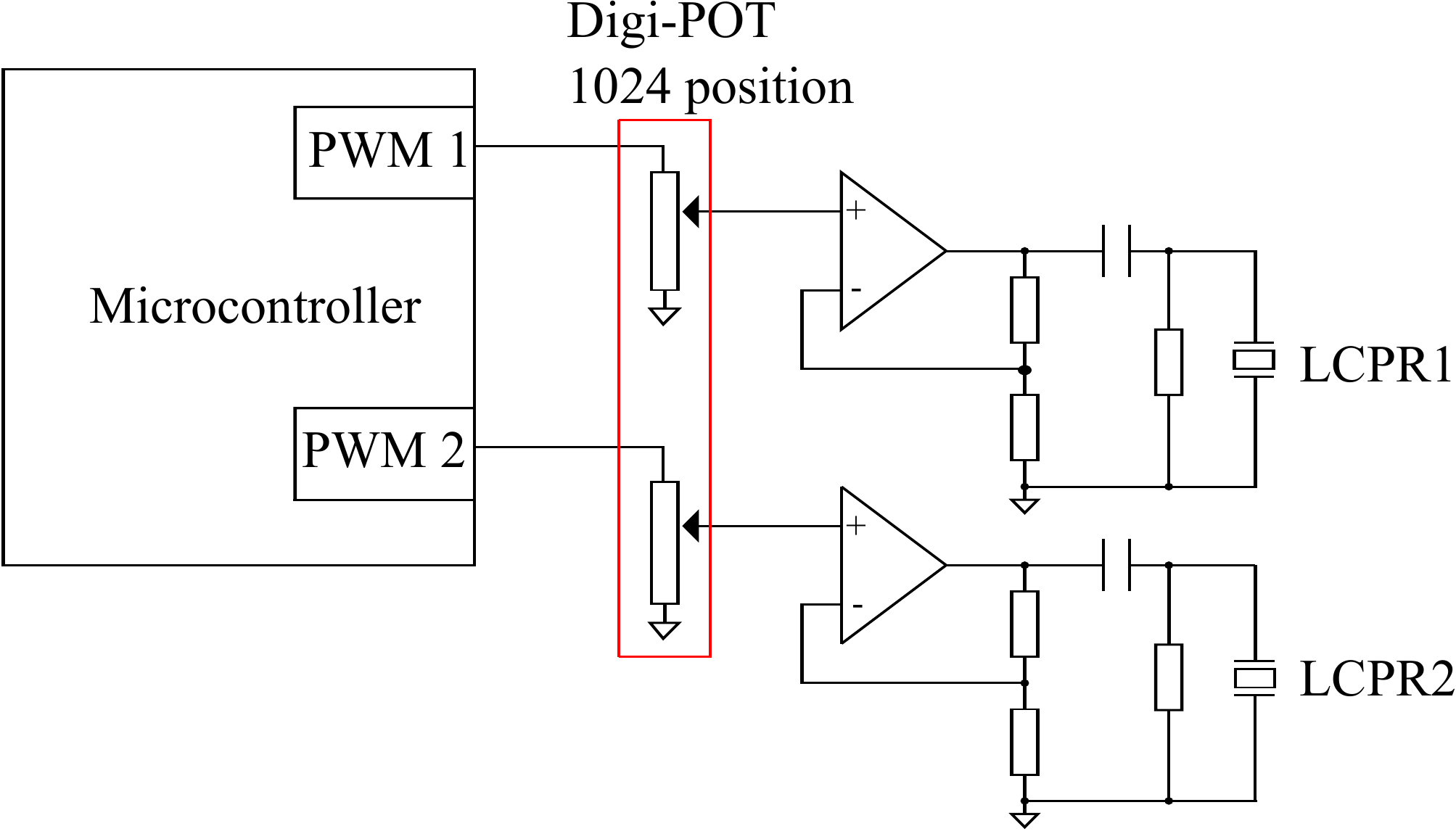}
  \vspace{0.8cm}
  \subcaption{{}\label{}}
  \label{fig:SPEQSElectronicsv121B1-img006}
  \end{minipage}
\hfil
\hspace{0.2cm}
 \begin{minipage}[c]{0.48\textwidth}
 \centering
  \vspace{0.3cm}
   \includegraphics[scale=0.85]{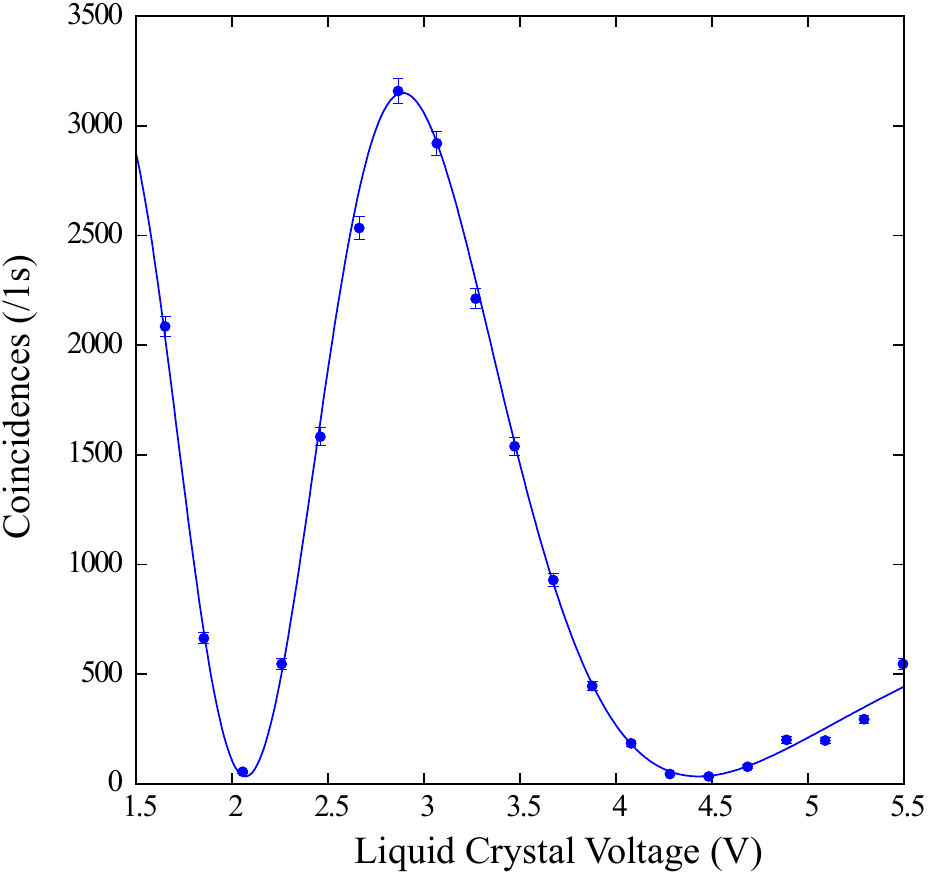}
  \subcaption{{}\label{}}
  \label{fig:SPEQSElectronicsv121B1-img006}
  \end{minipage}
  \caption{ (a) The polarization rotator module. To \ effectively measure polarization
correlations, it is necessary to rotate the polarization of the photons to be
analysed . In the SPEQS instrument, this rotation is achieved using liquid
crystal polarization rotators (LCPRs). The digital potentiometer (DigiPOT) is
used to apply the correct amplitude to the waveforms sent to the LCPRs. This
achieves inertial-free polarization rotation without using rotary stages and
waveplates.
   (b) Coincidence counts from the SPEQS instrument when the amplitude to one
LCPR  device is adjusted between \SI{1.5}{\volt} and \SI{5.5}{\volt}, corresponding to a half-wave
plate rotation of approximately \SI{90}{\degree}.}
\label{fig:SPEQSElectronicsv121B1-img006}
\end{figure*}

The GM-APD is supplied with bias voltage from a regulated source (Matsusada
TS-0.2P) capable of supply up to \SI{200}{\volt}. For the temperature variation within a nanosatellite, it was
determined that a \SI{40}{\volt} tuning range for the bias voltage (\SI{120}{\volt} to \SI{160}{\volt}) was
sufficient. It was also highly desirable to have the ability to tune the high
voltage output in steps of \SI{0.01}{\volt} in order to avoid over-shooting the optimal
operating voltage. However, the microcontroller cannot directly provide such a
high resolution over the entire \SI{40}{\volt} range.

To achieve this, a {\textquotedblleft}summing amplifier{\textquotedblright}
circuit was constructed as illustrated in
FIG.~\ref{fig:SPEQSElectronicsv121B1-img003}(b). A current DAC (iDAC) generates a
current in steps of \SI{0.5}{\micro\ampere} which is used to generate a variable voltage. \ A
standard DAC is programmed to produce an offset voltage. A summing operational
amplifier adds the offset voltage to the variable voltage and produces the final
control voltage. This setup achieves a continuous control of the voltage source
output with the required resolution.

\subsection{ Polarization Rotator Module}

To measure polarization correlations, it is necessary to analyse the
polarization of the photons over an entire basis, for example in the linear
basis. In a laboratory setup this is typically performed by rotating
wave-plates mechanically. Amongst other problems, this introduces torque on a
spacecraft, potentially interfering with its attitude control. To avoid this, we
have implemented an inertial-free polarization rotator based on liquid crystal
technology.

The liquid crystal polarization rotators (LCPRs) are customized for the target
wavelengths (\SI{860}{\nano\meter} and \SI{760}{\nano\meter}). The polarization rotation responds to the
amplitude of a DC-balanced square wave at \SI{3}{\kilo\hertz} generated by the PWM component
of the microcontroller. The PWM output is \SI{3.3}{\volt} logic fed into a dual digital
potentiometer. \ The wiper positions of the potentiometer can be adjusted so that the
final waveform amplitude sent to the LCPR can be adjusted in steps of
\SI{9}{\milli\volt}. FIG.~\ref{fig:SPEQSElectronicsv121B1-img006}(b) shows the high contrast
variation in the photon pair detection rate when one LCPR \ is supplied with
fixed voltage, and the other LCPR has its amplitude adjusted between \SI{1.5}{\volt} and \SI{5.5}{\volt}
(corresponding to a half wave plate rotation of approximately \SI{90}{\degree}).
This high contrast indicates that fine control over polarization rotation can be
achieved by the LCPRs.

\subsection{ Pump laser, thermal management and NAND-Flash storage}

\begin{figure}[ht]
\begin{center}
\includegraphics[width=0.495\textwidth]{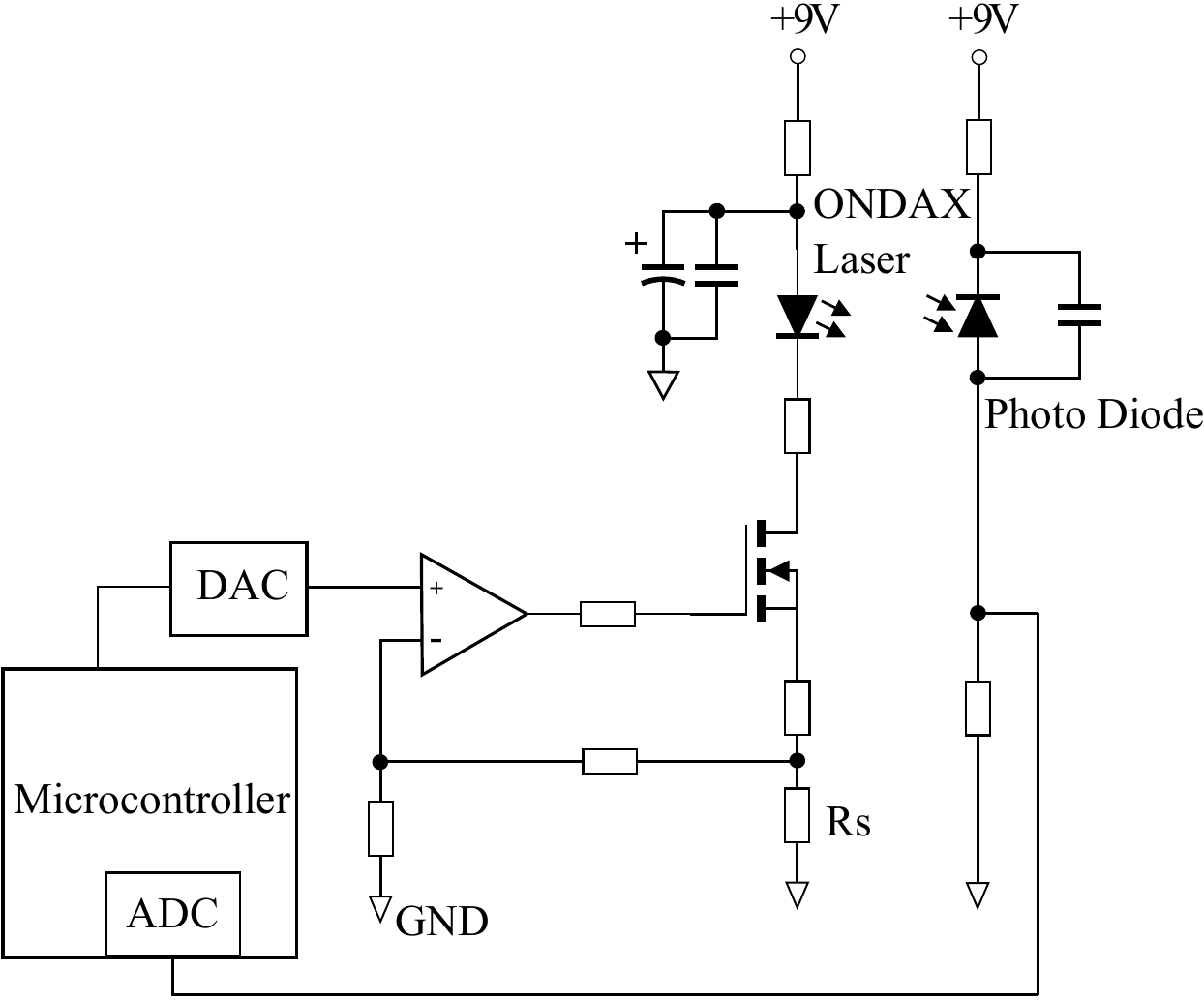}
\caption{The pump laser module. The pump laser is a \SI{405}{\nano\meter} GaN-based laser diode
and is always operated as a continuous-wave device. Two operational modes are
possible: constant current or constant power.}
\label{fig:SPEQSElectronicsv121B1-img007}
\end{center}
\end{figure}

The SPEQS optical unit contains a grating-stabilized GaN-based laser diode
emitting at \SI{405}{\nano\meter} (Ondax CP-405PLR40). The laser diode is always operated as a
continuous-wave device. For flexibility we implemented a module which can
operate the laser diode in constant current mode or in constant power mode. This
is illustrated in FIG.~\ref{fig:SPEQSElectronicsv121B1-img007}.

In constant power mode, an external Si PIN photodiode \ (Hamamatsu S5106)
configured in reversed bias monitors the power of the pump beam in the SPDC
process. As the SPDC efficiency is on the order of $10^{-11}$/mm, essentially
all the laser power is picked up by the photodiode. The microcontroller samples
the photodiode output periodically and adjusts the laser current, maintaining
the power at the set point.


In constant current mode, the voltage across a sense resistor (Rs) is used to
generate an error signal that is compared with a set value. A MOSFET switch is
used to control the laser current in steps of \SI{50}{\micro\ampere}. For a smaller current step
size, the value of the sense resistor can be increased.

\begin{figure}[h]
\begin{center}
\includegraphics[width=0.485\textwidth]{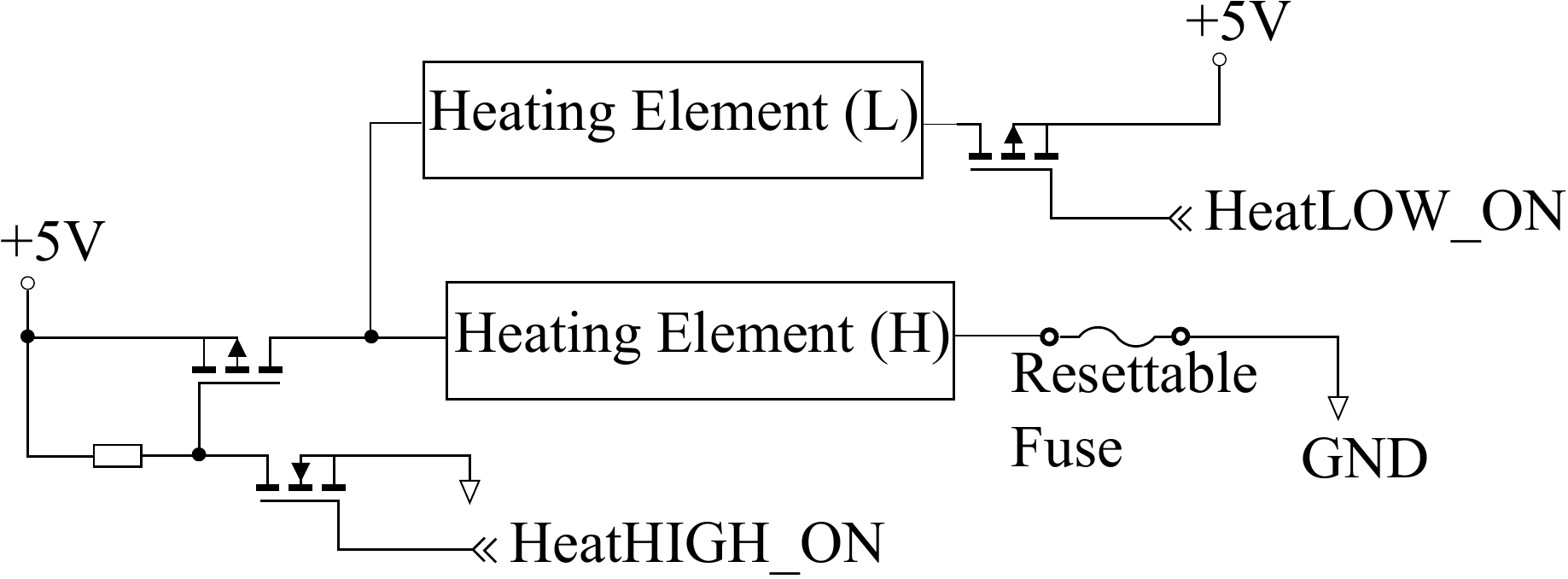}
\caption{ The heating element module for maintaining the temperature of the
optical apparatus.}
\label{fig:SPEQSElectronicsv121B1-img008}
\end{center}
\end{figure}

The optical assembly is always constructed at approximately \SI{22}{\celsius} and
can operate on either side of this up to a range of \SI{10}{\celsius}. However,
the nanosatellite thermal environment can oscillate between \SI{-15}{\celsius} to
\SI{12}{\celsius} \cite{Kataoka2010_1}, hence it is desirable to
stabilize the temperature of the optical assembly. A heating module operates in two modes:
{\textquotedblleft}high{\textquotedblright} and
{\textquotedblleft}low{\textquotedblright}. The implementation is illustrated in
FIG.~\ref{fig:SPEQSElectronicsv121B1-img008}.

In the {\textquotedblleft}high{\textquotedblright} mode, a heating element (H)
draws \SI{2.5}{\watt} of power to rapidly raise the temperature of the optical assembly to
within operating range. This mode will be used when the SPEQS instrument is
powered on. Repeated testing under vacuum has shown that this mode can raise the
optical assembly temperature by approximately \SI{1}{\celsius} per minute.


In the {\textquotedblleft}low{\textquotedblright} mode, a secondary heating
element (L) is added in series to H, drawing a maximum of \SI{0.4}{\watt}. This is
performed by selectively turning on a MOSFET switch. Testing has demonstrated
that in conjunction with the heat generated by the pump laser, the temperature
of the optical assembly can be maintained within the operating range.

\begin{figure*}[t]
\centering
  \begin{minipage}[t]{0.48\textwidth}
  \centering
  \hspace{-0.9cm}
  \includegraphics[scale=0.3,valign=t]{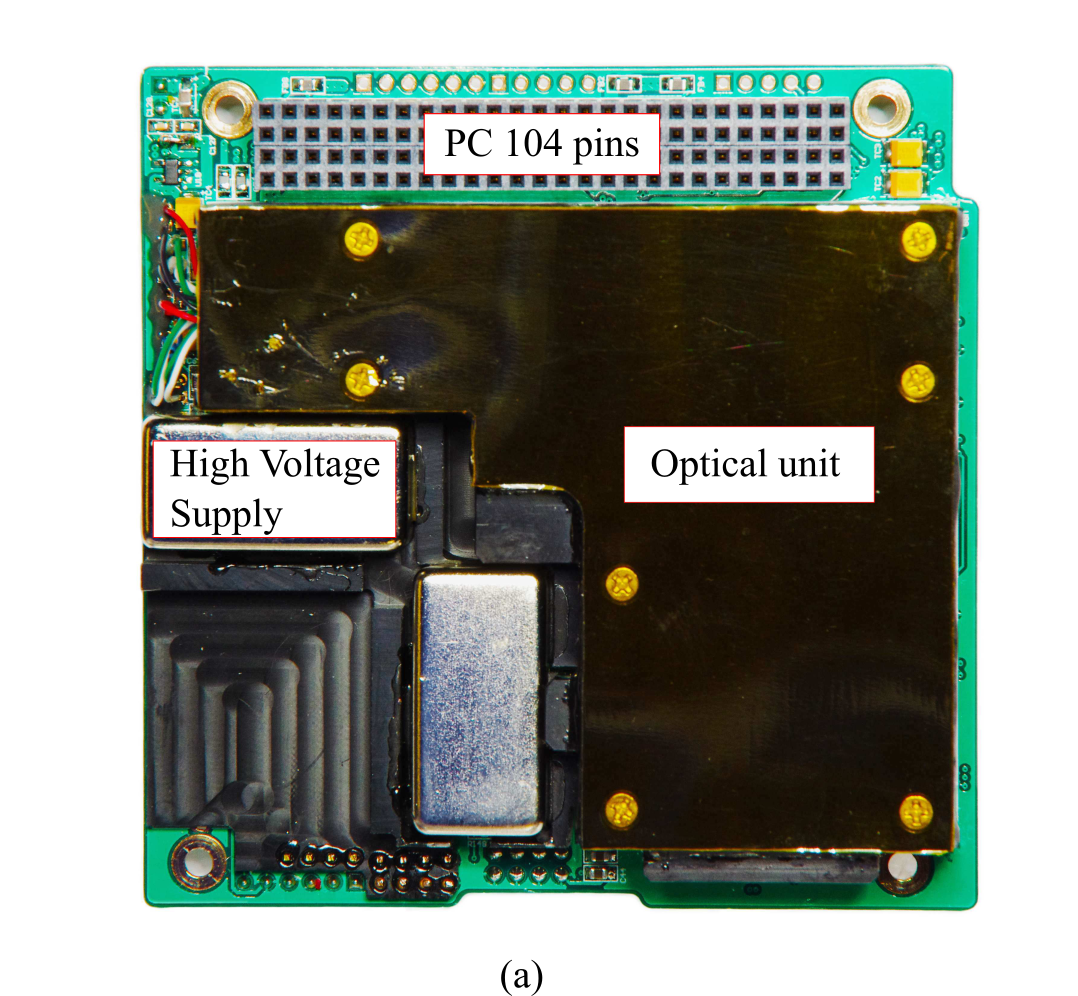}
  \vspace{0.2cm}
  \label{fig:SPEQSElectronicsv121B1-img009}
  \end{minipage}
\hfil
 \begin{minipage}[t]{0.48\textwidth}
 \centering
  \vspace{-0.5cm}
  \includegraphics[scale=0.33,valign=t]{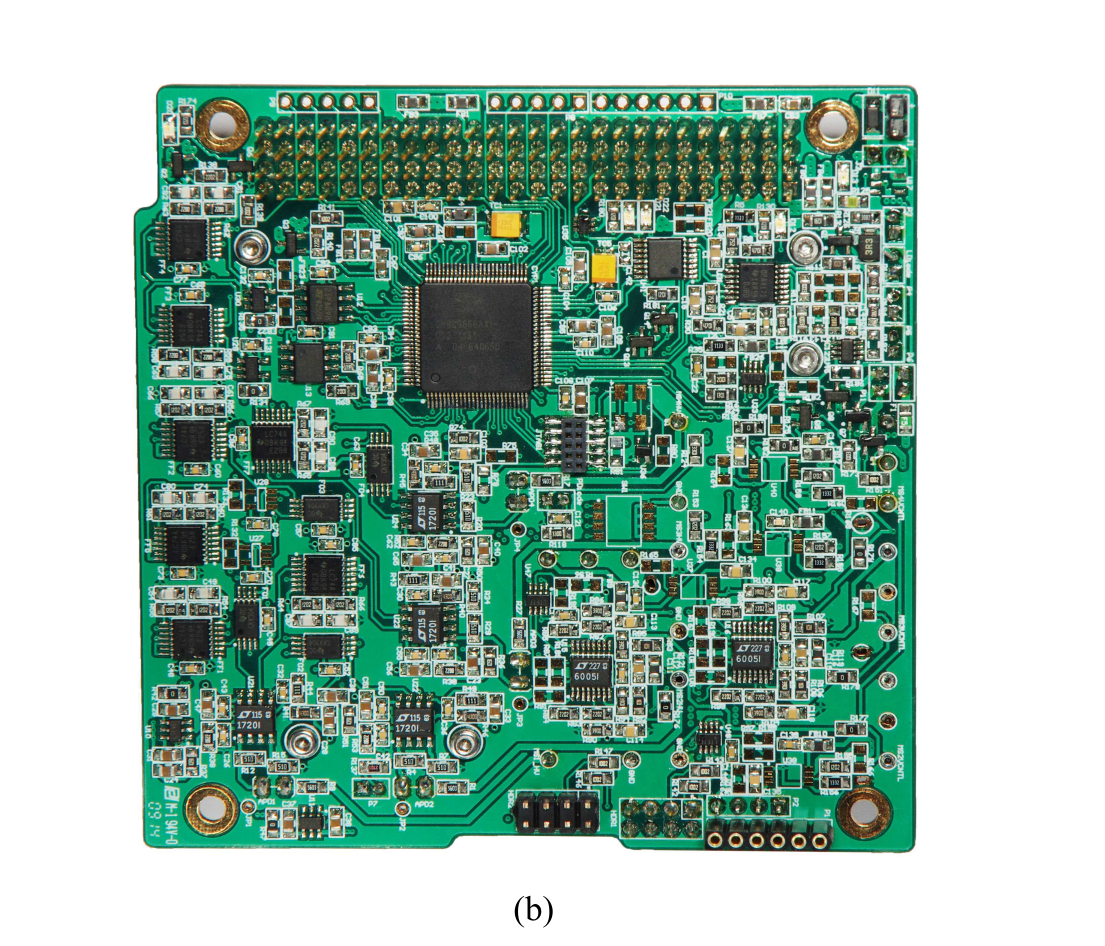}
 \vspace{-0.4cm}
   \hspace{-0.4cm}
  \label{fig:SPEQSElectronicsv121B1-img009}
  \end{minipage}
 \vspace{-0.1cm}
  \caption{ (a) The optical assembly is held within a black-anodized aluminium
box that is attached to one side of the electronics platform. Most of the
control circuitry is on the reverse side. Mechanical interfacing with the
spacecraft is done via four mounting holes at the corners of the printed circuit
board, as well as through the PC-104 pins. (b) The complete electronic logic in a CubeSat compatible form-factor.}
\label{fig:SPEQSElectronicsv121B1-img009}
\end{figure*}

Experiment data is stored on a 8-Mbit SPI-based NAND flash memory device with 16
sectors. Each sector of memory is 65536 bytes. To alleviate the
possibility of corruption by radiation each set of record data is stored
redundantly in two different sectors of the memory at \SI{1.3}{\second} intervals. Each
frame of data is 32 bytes, and each page of memory is 256 bytes, allowing 7
frames of data to be written into each page. With this configuration it is
possible to store up to \SI{30}{\minute} of data in each sector. 

\section{ Discussion and Conclusion}

The form-factor of the electronics platform is designed to conform to the
CubeSat design specification \cite{TheCubeSatProgram2009}. The printed circuit
board housing the electronics measures \SI{95}{\milli\meter} x \SI{95}{\milli\meter}. With the optical assembly
mounted, the overall height of the package is 38~mm, and the entire instrument
mass is less than \SI{220}{\gram}. The electronics platform incorporates a stackable PC-104
bus for power and signal connections with the rest of the spacecraft. The
electronics platform is also relatively efficient in power consumption, and in
experiment mode (running a pump laser and two high-efficiency GM-APDs), consumes
less than \SI{1.5}{\watt} (see TABLE \ref{table:table_1}). We anticipate future power
savings when a large number of the electronic operations are moved into
integrated devices such as complex programmable logic devices (CPLDs) or field
programmable gate arrays (FPGAs). This migration into software-defined circuits
could further reduce design complexity while improving system robustness. 

%
%

\begin{table}[h]
\renewcommand{\arraystretch}{1.3}
\caption{\sc{The observed power consumption for each module in the electronics
platform. *Thermal Management (High) only operates prior to start of experiment
and data collect.}}
\label{table:table_1}
\centering
\begin{tabular}{c|c}
\hline
\bfseries Electronics Platform Sub-Systems & \bfseries Power Consumption (W)\\
\hline
Thermal Management (High)*  & 2.5\\
\hline
Thermal Management (Low) & 0.4\\
\hline
GM-APD Control & 0.24\\
\hline
Pump Laser Management & 0.45\\
\hline
LCD Polarisation Rotator & 0.1\\
\hline
PsoC3 Operation @ 24MHz (Normal)\\
-Computation, Data Storage, \\
USART Communications  & 0.3\\
\hline
PsoC3 Operation (Standby) & 0.1\\
\hline
\end{tabular}
\end{table}

The electronics platform has been tested successfully in radiation
\cite{Tan2013}, thermal-vacuum and vibration environments to simulate launch and
operation in space. Further testing in a near-space environment using a high
altitude balloon \cite{Tang2014} was also successful. The first attempt at
putting a SPEQS experiment into orbit occurred in 2014 when the
instrument was integrated onto the GomX-2 satellite (see Appendix A) that was lost in a launch
vehicle failure. Because the SPEQS experiment has followed the CubeSat standard,
it has been accepted on other CubeSat missions. Performance data from low Earth
orbit is expected to be available in early 2016.

We have presented the main blocks of an electronics platform that can support
the operation of an entangled photon source on a 1U CubeSat. The concept of the
integrated electronics platform for supporting space-based quantum
communications has been demonstrated, and the design will be utilised in future
SPEQS-based missions. 

\section*{Acknowledgment}
During the development of this platform, C. Cheng and Tan Y. C. were supported
by the DSO-CQT project on quantum sensors. Both of them are currently supported
by the National Research Foundation project NRF-CRP12-2013-02. The authors
thanks R. Bedington for assistance with the manuscript.

\newpage

\appendices
\section{}
\begin{figure}[!h]
\begin{center}
 \includegraphics[scale=0.095]{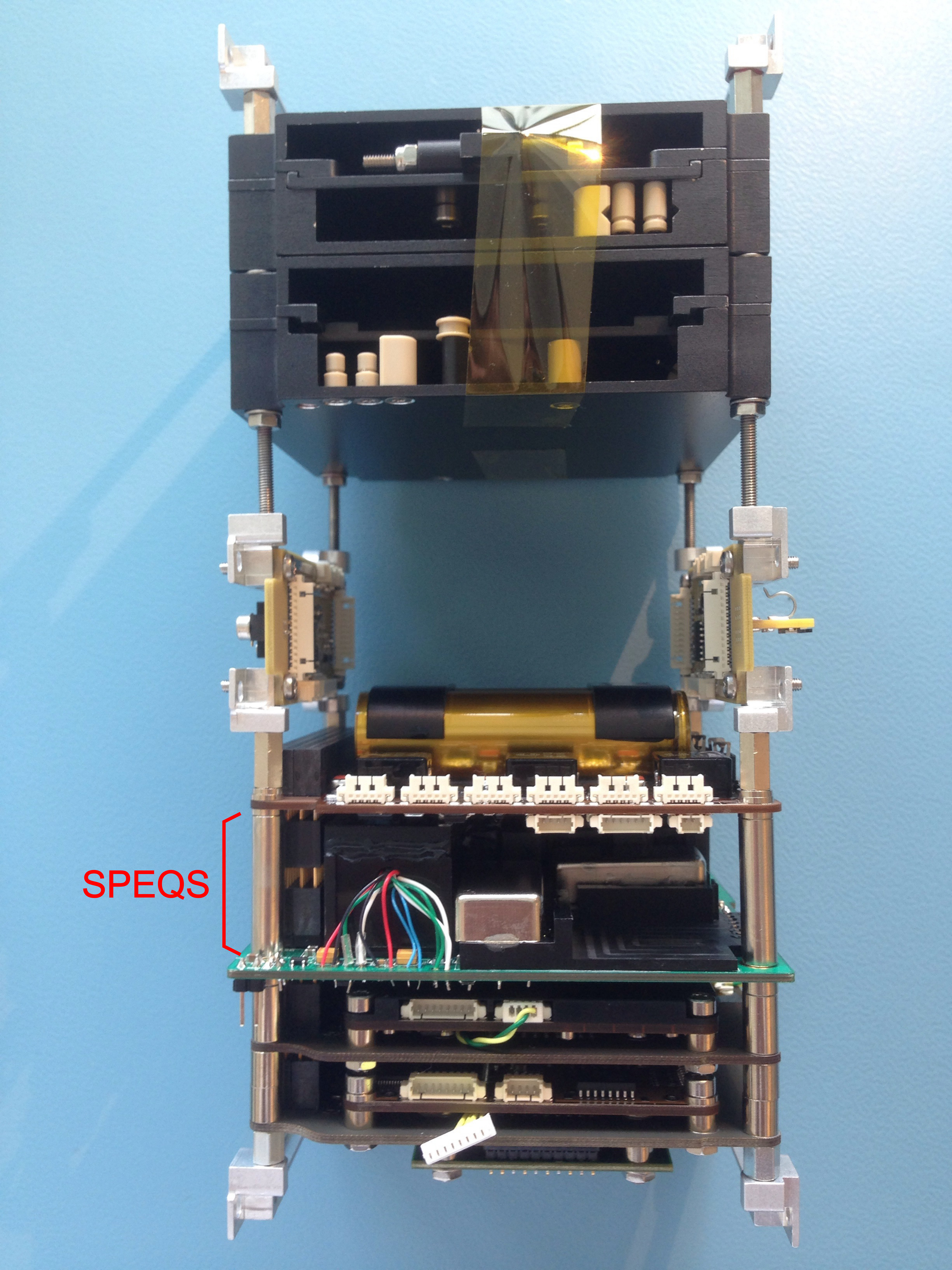}
\caption{The SPEQS experiment during installation into the GomX-2
satellite. This picture is published with permission from GomSpace ApS.}
\label{fig:CS5back1}
\end{center}
\end{figure}

%
%
%
%
%
%
%


%
%

\ifCLASSOPTIONcaptionsoff
  \newpage
\fi

\newpage



\bibliographystyle{IEEEtran}
\bibliography{library}
\end{document}